\documentclass{mn2e}
\usepackage{psfig}
\usepackage{mnras_cite}
\newcommand{\gsim}{\;\rlap{\lower 2.5pt
 \hbox{$\sim$}}\raise 1.5pt\hbox{$>$}\;}
\newcommand{\lsim}{\;\rlap{\lower 2.5pt
   \hbox{$\sim$}}\raise 1.5pt\hbox{$<$}\;}
\newcommand{\be}{\begin{equation}}
\newcommand{\beq}{\begin{equation}}
\newcommand{\ba}{\begin{eqnarray}}
\newcommand{\ee}{\end{equation}}
\newcommand{\eeq}{\end{equation}}
\newcommand{\ea}{\end{eqnarray}}

\newcommand{\rad}{r}

\newcommand{\wjm}{\left(
                           \begin{array}{ccc}
         l_1 & l_2  & l_3  \\
         m_1 & m_2  & m_3
                           \end{array}
                   \right)}

\newcommand{\bi}{B_{l_1 l_2 l_3}}

\newcommand{\Ylm}[1]{Y_{l_#1}^{m_#1}}

\newcommand{\alm}[1]{a_{l_#1 m_#1}}

\newcommand{\bea}{\begin{eqnarray}}
\newcommand{\eea}{\end{eqnarray}}
\newcommand{\bean}{\begin{eqnarray*}}
\newcommand{\eean}{\end{eqnarray*}}

\newcommand{\bfx}{{\bf x}}
\newcommand{\br}{{\bf r}}
\newcommand{\bk}{{\bf k}}
\newcommand{\bn}{{\bf \hat{n}}}
\newcommand{\bl}{{\bf \hat{l}}}
\newcommand{\bm}{{\bf \hat{m}}}
\newcommand{\bfn}{{\bf \hat{n}}}
\newcommand{\bfk}{{\bf \hat{k}}}

\newcommand{\nn}{\nonumber\\}

\begin{document}

\title[Non-Gaussianities in 21 cm Anisotropies]{Large-Scale Non-Gaussianities in the 21 cm Background Anisotropies 
From the Era of Reionization}
\author[Cooray]{Asantha Cooray \\
Theoretical Astrophysics, MS 130-33, California Institute of Technology, Pasadena, CA 91125.\\
Dept. of Physics and Astronomy, 4129  Frederick Reines Hall, University of California,   Irvine, CA 92697.\\
 E-mail: asante@caltech.edu}

\maketitle

\begin{abstract}
The brightness temperature fluctuations in the 21 cm background related to neutral Hydrogen distribution
provide a probe of physics related to the era of reionization when the intergalactic medium changed from a
completely neutral to a partially ionized one. We formulate statistics of 21 cm
brightness temperature anisotropies in terms of the angular power
spectrum, the bispectrum, and the trispectrum. Using the trispectrum, we 
estimate the covariance related to the power spectrum measurements and show that correlations resulting from
non-Gaussianities are below a percent, at most.
While all-sky observations of the 21 cm background at arcminute-scale resolution can be used to measure the
bispectrum with a cumulative signal-to-noise ratio of order a few ten, in the presence of foregrounds
and instrumental noise related to first-generation interferometers, the measurement is unlikely to be feasible.
For most purposes, non-Gaussianities in 21 cm fluctuations can be ignored and the distribution can be
described with Gaussian statistics. Since 21 cm fluctuations are significantly contaminated by foregrounds, such
as galactic synchrotron or low-frequency radio point sources, the lack of a significant non-Gaussianity in the
signal suggests that any significant detection of a non-Gaussianity can be due to foregrounds.
Similarly, in addition to frequency information that is now proposed to separate 21 cm fluctuations from foregrounds,
if the non-Gaussian structure of foregrounds is a priori known, one can potentially use this additional information to
further reduce the confusion. 
\end{abstract}


\section{Introduction}

The 21 cm radio background related to spin-flip transition is a well-known probe of the neutral
Hydrogen density content prior to and during the era of reionization
\cite{Scott90,madau97,zaldarriaga04,morales03}. 
This background has the advantage that by selecting the observed frequency of the redshifted line emission,
one can directly view the neutral Hydrogen distribution of the Universe at a given redshift. With maps
of the neutral Hydrogen content, as a function of the redshift, one can obtain statistics related to the reionization process,
especially during the transition from a fully neutral medium to an ionized one \cite{BarLoe01}. Similarly, prior to the epoch
of reionization when the Universe is fully neutral, 21 cm fluctuations provide a probe of the primordial 
density perturbations \cite{loeb04,bharadwaj04}.

In addition to 21 cm data,  information related to the era of reionization  can be obtained with large angular scale cosmic microwave
background (CMB) polarization observations.  With CMB anisotropies, one is studying the presence of electrons through various Compton-scattering
related processes where modifications are introduced to the temperature and polarization fluctuations. 
Unlike 21 cm observations, unfortunately, CMB data only allow a measurement of the
integrated column density of electrons \cite{zaldarriaga97a} with no detailed information on the exact reionization history of the universe
\cite{hu03,kaplinghat03}. As has been studied recently, Lyman-$\alpha$ optical depth from Gunn-Peterson
troughs \cite{gunn65} towards $z \sim 6$ quasars in the Sloan Digital Sky Survey \cite{fan01} provides information 
related to the end of reionization process when the Universe is almost fully ionized.
While recent measurements from the WMAP mission \cite{bennett03} suggest an optical depth to electron scattering of 
0.17 $\pm$ 0.04 \cite{Kogut03}, it can be explained with a variety of reionization models that also take into account the
ionizing fraction estimates from $z \sim$ 6 quasars \cite{cen03,Ciardi03,chen03,gnedin04b,wyithe03,haiman03}. 

With 21 cm observations, one can potentially reconstruct the exact reionization history and determine, especially, if the reionization 
process was either instantaneous or lengthy and complex
as implied by current CMB and Sloan data. Moreover, since observable effects in CMB data depend only
on the ionized content while  21 cm fluctuations come from inhomogeneities in the neutral 
distribution, the combination is expected to allow additional information related to the era of reionization \cite{Coo04a,Coo04b}.
Detailed analytical and numerical studies related to the physics associated with the 21 cm background are now motivated by
progress in the experimental side involving plans 
for a variety of low-frequency radio experiments such as PAST\footnote{http://astrophysics.phys.cmu.edu/~jbp}
\cite{pen04}, Low Frequency
Array (LOFAR)\footnote{http://www.lofar.org}, and Square Kilometer Array (SKA)\footnote{http://www.skatelescope.org}.

While the scientific motivation to study 21 cm fluctuations during the era of reionization is 
strong, there are various challenges to overcome. These include the removal of foreground radio emission, such as due to
synchrotron emission from the Milky Way \cite{shaver99}, low frequency radio point sources
\cite{dimatteo02} and free-free emission from free electrons in
the intergalactic medium \cite{oh99,cooray04}. Contrary to initial suggestions
that 21 cm background studies may be impossible due to
the large number of contaminating sources \cite{oh03}, with brightness fluctuations much higher than expected for neutral Hydrogen at 
high redshifts, recent suggestions based on multifrequency analysis indicate that foreground contamination can be 
reduced given the smoothness of these confusions in frequency space \cite{shaver99,dimatteo02,zaldarriaga04,santos04}. 

In current studies, the statistics related to 21 cm fluctuations have been mostly restricted to the angular power spectrum of anisotropies.
This is primarily motivated by the expected Gaussian nature of these fluctuations. It is useful to consider if there are
non-Gaussian signals in the 21 cm background anisotropies and, if any exists, whether these signals can be measured with
planned low-frequency instruments.  Here, we discuss the role of non-Gaussianities in the 21 cm background.
For this purpose, we describe spatial fluctuations in the 21 cm background during the era of reionization  as due to
a combination of spatial fluctuations in the underlying neutral Hydrogen density, peculiar velocity of neutral gas, and the fraction of
ionized gas relative to the total gas density. 
The non-Gaussian signals are generated by the non-linear coupling of fluctuations in either the
neutral density and peculiar velocity  or neutral density and ionized-fraction fields. 

We model the trispectrum, the Fourier analog of the four-point correlation  function, of 21 cm anisotropies and
explore how these non-Gaussianities affect the power spectrum measurement in terms of its full covariance.
The trispectrum produces an additional source of noise beyond the standard cosmic variance under Gaussian statistics and correlates
power spectrum measurements between different multipoles \cite{sco99,cooray01}. 
As we find, in the case of the 21 cm background, these correlations   are at the level of at most a percent
suggesting that non-Gaussianities are not likely to be a major source of complication when making a power spectrum measurement. 
We also estimate the bispectrum, the three-point correlation in Fourier space,
of 21 cm anisotropies. While all-sky observations with no significant instrumental noise
can be used to measure the bispectrum with signal-to-noise ratios of a few ten, for more practical observations with
the first-generation of low-frequency radio interferometers, the signal-to-noise ratio is below a unity. In general,
21 cm background anisotropies can be described with
Gaussian statistics and measurements can be focused primarily on the power spectrum. In our calculations, we mainly focus
on the large physical, and thus large angular, scale fluctuations of the 21 cm background. Non-linear effects related to
the neutral gas distribution and velocities can lead to additional non-Gaussian signals, but such effects are
best captured in numerical simulations of the reionization process. These include the effects associated with
finite size of reionized regions (c.f., Bharadwaj \& Pandey 2004), and we leave these and other issues to be discussed in future works.
For first-generation observations, such effects are unlikely to be
important given restrictions on the angular resolution of observations. The lack of a significant non-Gaussianity
at large angular scales, at least, has the advantage that if any significant non-Gaussianity is detected, it can
be ascribed to  foreground sources. Thus, in addition to smooth behavior in frequency space, the non-Gaussian structure of
foregrounds, such as those related to Galactic synchrotron, can be used to further reduce the confusion.

The paper is organized as follows: in \S \ref{21cm}, we discuss
the 21 cm signal anisotropy, the trispectrum in terms of the
power spectrum covariance, and the bispectrum. While we only consider the non-Gaussian structure of the 21 cm background,
the lack of a significant non-Gaussianity in the signal can be potentially used to separate nearby galactic foregrounds
which are expected to be highly non-Gaussian. We discuss these and other possibilities in
\S \ref{discussion}. Throughout the paper, we make use of the WMAP-favored $\Lambda$CDM cosmological model \cite{spergel03}.
                                                                                                                            
\section{The 21cm signal}
\label{21cm}

\begin{figure}
\centerline{\psfig{file=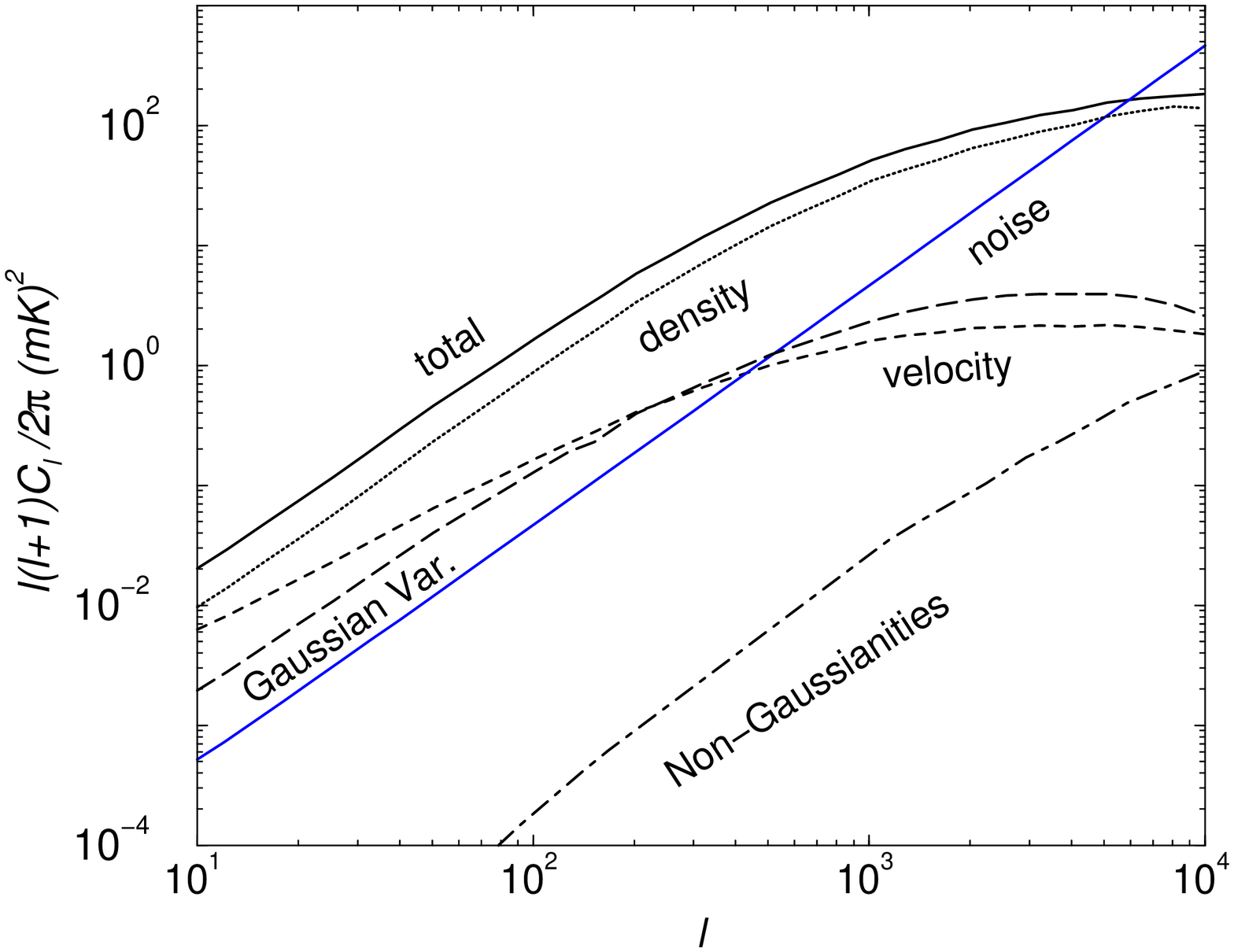,width=3.5in,angle=-90}}
\centerline{\psfig{file=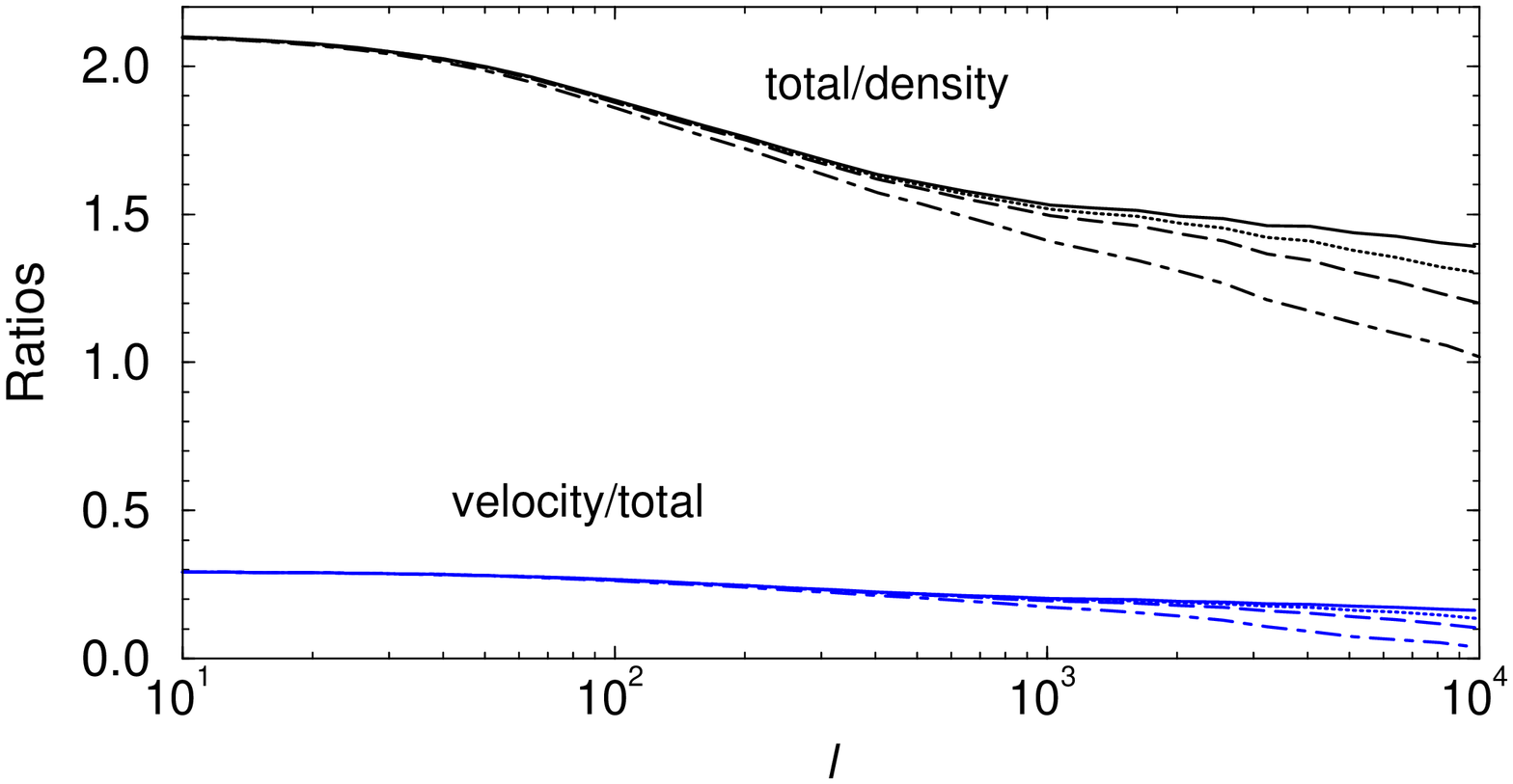,width=3.5in,angle=-90}}
\caption{{\it Top:} The power spectrum for the 21cm signal at z=10.0 and assuming a bandwidth of
$\delta \nu=1$MHz. Units are in mK$^2$. The calculation makes use of a
reionization model from Santos, Cooray \& Knox (2004), though similar results are obtained for
slightly varying models from Santos et al. (2003). Here, we show the total contribution to the power spectrum (solid line),
as well as the power spectrum related to neutral density (dotted line) and peculiar velocity (dashed line) fluctuations alone.
We also show the Gaussian cosmic variance (long-dashed line) and an additional contribution to the power spectrum variance
related to non-Gaussian trispectrum (dot-dashed line; see \S 2.1 for details). The solid line labeled 'noise' shows
an estimated instrumental noise curve related to 21 cm observations with the SKA following Zaldarriaga, Furlanetto \& Hernquist (2004).
{\it Bottom:} The ratio of various contributions to the 21 cm anisotropy power spectrum. In the top curves, we show the ratio of
total contribution to that related to density fluctuations alone, while the bottom set of curves are the ratio of peculiar
velocity contribution to that of the total power spectrum. In each of these two sets, the four curves, from top to bottom, are
for observational bandwidths of 0.1, 0.5, 1.0 and 2.5 MHz, respectively.}
\label{fig:power}
\end{figure}

The 21 cm surface brightness temperature observations are primarily observed as a change in the intensity of
CMB due to line emission and absorption. The change in brightness temperature,
$T_{21}(\nu)$, compared to the CMB at an observed frequency $\nu$ is then
\bea
T_{21}(\nu) & \approx & \frac{T_S - T_{\rm CMB}}{1+z} \, \tau 
\label{eq:dtb} 
\eea
where $T_S$ is the spin temperature of the IGM, $z$ is the
redshift corresponding to the frequency of observation ($1+z=\nu_{21}/\nu$, with 
$\nu_{21} = 1420$ MHz) 
and $T_{\rm CMB} = 2.73 (1+z) K$ is the CMB temperature at redshift $z$.

In order to simplify the calculation related to  21 cm fluctuations, we assume that $T_s\gg T_{CMB}$, in a scenario where the gas density field
in the intergalactic medium is heated as part of the reionization and structure formation processes.  These include
shocks associated with the formation of virialized dark matter halos where first stars are eventually found, and whose UV photons 
subsequently reionize the universe, and X-rays from first supernovae \cite{venkatesan01,chen04}.
At sufficiently high $z$, prior to significant reionization and heating, we expect $T_s < T_{\rm CMB}$,
and the HI density content is seen as an absorption in 21 cm with respect to the CMB \cite{madau97}.
The statistics of 21 cm fluctuations related to such an early epoch can be used as a probe of primordial fluctuations \cite{loeb04,bharadwaj04}.

The optical depth, $\tau$, of this patch in the hyperfine transition \cite{field59} is given 
in the limit of $k_B T_s >> h \nu_{21}$ by
\bea
\tau & = & \frac{ 3 c^3 \hbar A_{10} \, n_{\rm HI}}{16 
k \nu_{21}^2 \, T_S \, H(z) } 
\label{eq:tauigm} \\
\, & \approx & 8.6 \times 10^{-3} \left(1+\delta_g - \frac{1+z}{H(z)}\frac{\partial v}{\partial r}\right) x_H \left[
\frac{T_{\rm CMB}(z)}{T_S} \right] 
\nonumber \\
\, & \, & \times
\left( \frac{\Omega_b h^2}{0.02} \right) 
\left[ \left(\frac{0.15}{\Omega_m h^2} \right) \, \left(
\frac{1+z}{10} \right) \right]^{1/2} \left( \frac{h}{0.7}
\right)^{-1}, 
\nonumber
\eea
where $A_{10}$ is the spontaneous emission coefficient for the transition ($2.85 \times 10^{-15}$ s$^{-1}$),
$n_{\rm HI}$ is the neutral Hydrogen density, and 
$v$ is the peculiar velocity of the neutral gas distribution when $r$ is the comoving radial distance and $H(z)$ is the expansion rate
at a redshift of $z$. The neutral density be expressed as
$n_{\rm HI}=x_H\bar{n}_g(1+\delta_g)$, when $\bar{n}_g$ is the mean number density of cosmic baryons, with a
spatially varying overdensity $\delta_g$ and
$x_H$ is the fraction of neutral Hydrogen ($x_H= 1-x_e$ where $x_e$ is the fraction of
free electrons). 
We refer the reader to  Zaldarriaga, Furlanetto \& Hernquist (2004) and Santos, Cooray \& Knox (2004) for further details.
Note, however, that prior calculations related to the 21 cm anisotropies ignored both peculiar velocities and ionizing
fraction fluctuations (e.g., Zaldarriaga, Furlanetto \& Hernquist 2004), or peculiar velocities (Santos, Cooray \& Knox 2004).

With $T_s >> T_{\rm CMB}$, the 21 cm signal will be observed in emission with respect to the background CMB.
Writing the brightness temperature as:
\be
T_{21}(z)=c\,\left(1+\delta_g - \frac{1+z}{H(z)}\frac{\partial v}{\partial r}\right)\left(1-\bar{x}_e-\bar{x}_e\delta_x\right),
\ee
where 
\be
c\approx 23\left( \frac{\Omega_b h^2}{0.02} \right) \nonumber \\
\left[ \left(\frac{0.15}{\Omega_m h^2} \right) \, \left(
\frac{1+z}{10} \right) \right]^{1/2} \left( \frac{h}{0.7}
\right)^{-1} {\rm mK} \, ,
\ee
and $\delta_x$ is the perturbation in the ionization fraction ($\delta_x\equiv {x_e-\bar{x}_e\over\bar{x}_e}$).
Here and above, the spatial dependent quantities are evaluated at the position $\bfx = r \bfn$ and for
simplicity in notation we have dropped this dependence.

The measured brightness temperature corresponds to a convolution of the intrinsic
brightness with some response function $W_\nu(r)$ that characterizes the bandwidth of the
experiment:
\be
T(\bn,\nu_0)=\int dr W_{\nu_0}(r) T_{21}(\bn r,r),
\ee
where $\bn$ is the direction of observation and $r$ corresponds 
to the observed frequency $\nu$.                                                                                            
We can now determine the angular power spectrum of $T(\bn,\nu_0)$ which is related to the 3-d
power spectrum of $T_{21}(\bn,r)$, defined by
\be
\label{3dpower}
\langle\tilde{T}_{21}(\bk,\nu_1)\tilde{T}_{21}(\bk',\nu_2)\rangle=
(2\pi)^3\delta^D(\bk+\bk') P_{21}(k,\nu_1,\nu_2),
\ee
where $T_{21}(\hat{n},r)\equiv T_{21}(\br,\nu)=
\int {d^3k\over (2\pi)^3} \tilde{T}_{21}(\bk,\nu) e^{i\bk\cdot\br}$.

Making use of the spherical harmonic moment of the 21 cm fluctuations, at a frequency $\nu_0$,
\begin{equation}
a_{l m}(\nu_0) = \int d\bn Y^*_{l m}(\bn) T(\bn,\nu_0) \, ,
\end{equation}
we define the angular power spectrum as \cite{kaiser92}
\bea
\label{cl21}
& &\langle a_{l m}(\nu_1) a^{s*}_{l m}(\nu_2)\rangle=C_l(\nu_1,\nu_2) \, .
\eea

To calculate this power spectrum, first, we write Fourier space fluctuations in the brightness temperature as
\bea
T_{21}(\bk, z)&=&c\,[\delta_g(\bk)(1-\bar{x}_e) + \mu^2 \delta_v(\bk) (1-\bar{x}_e) - \bar{x}_e\delta_x(\bk)  \nonumber  \\ 
&& \quad - \bar{x}_e \int \frac{d^3\bk'}{(2\pi)^3} \delta_x(\bk') \delta_g(\bk-\bk') \nonumber \\
&& \quad - \bar{x}_e \mu'^2  \int \frac{d^3\bk'}{(2\pi)^3} 
\delta_x(\bk')  \delta_v(\bk-\bk')]
\eea
where $\delta_v$ is the velocity divergence, $\mu = \bfn \cdot \bfk$, and $\mu' = \bfn \cdot (\bfk-\bfk')$. 
Here, we concentrate primarily on the
large scale velocity fluctuations and at these scales, we can use linear theory to write $\delta_v(\bk) = f(\Omega_m) \delta^{\rm lin}(\bk)$
where $f(\Omega_m) \equiv \partial \ln G/ \partial \ln a \approx \Omega_m^{0.6} +1/70 [1- 1/2 \Omega_m(1+\Omega_m)]$, where $G$ is the
growth factor related to evolution of density perturbations with the
approximation valid in a flat cosmology with a cosmological constant.

To calculate the  power spectrum of 21 cm fluctuations, we ignore terms which are second order and higher and involving the
convolution between density and ionization-fraction and velocity and ionization-fraction fluctuations.
We make use of the Rayleigh expansion for the plane wave given by
\begin{equation}
e^{i\bk\cdot\br}= 4\pi \sum_{l m} i^{l} j_l(k r) Y_l^{m}(\bn) Y_l^{m*}(\hat{\bk}) \, ,
\end{equation}
and separately consider fluctuations in the velocity and density fields separately. To simplify the calculation associated with the
velocity fluctuations, we also use
\begin{equation}
\bfn \cdot \bfk = \sum_{m} \frac{4 \pi}{3} Y_1^{m}(\bn) Y_1^{m*}(\hat{\bk}) \, .
\end{equation}

After straightforward, but tedious, algebra, we can write the power spectrum as
\bea
\label{cl22}
&&C_l(\nu_1,\nu_2)={2\over\pi}\int k^2 dk\, \nn
&\times&\Big\{ P_{aa}(k,\nu_1,\nu_2) I_l^{\nu_1}(k) I_l^{\nu_2}(k) + P_{bb}(k,\nu_1,\nu_2) J_l^{\nu_1}(k) J_l^{\nu_2}(k) \nn
&+&P_{ab}(k,\nu_1,\nu_2) [I_l^{\nu_1}(k) J_l^{\nu_2}(k) + J_l^{\nu_1}(k) I_l^{\nu_2}(k)] \Big\}
\eea
where
\begin{eqnarray}
I_l^{\nu}(k) &=& \int dr W_{\nu}(r) j_l(k r) \nonumber \\
J_l^{\nu}(k) &=& \int dr W_{\nu}(r) j_l''(k r) \, ,
\label{eqn:J}
\end{eqnarray}
with the spherical Bessel function given by $j_l(k r)$. In deriving Eq.~\ref{cl22}, we have ignored terms which are second order in $\delta$.
In Eq.~\ref{eqn:J}, the double derivative of the spherical Bessel function is with respect to its argument.

The corresponding three-dimensional power spectra are 
\bea
\label{p21}
P_{aa}(k,z)&=&c^2\left[(1-\bar{x}_e)^2 P_{gg}(k,z)+\bar{x}_e^2 P_{\delta_x \delta_x}(k,z)\right. \\
&-& \left. 2 P_{g \delta_x}(k,z)\bar{x}_e(1-\bar{x}_e)\right] \, ,\nn
P_{bb}(k,z)&=&c^2\left[(1-\bar{x}_e)^2 P_{\delta_v \delta_v}(k,z)\right] \, ,\nn
P_{ab}(k,z)&=&c^2\left[(1-\bar{x}_e)^2 P_{g \delta_v}(k,z) - P_{\delta_v \delta_x}(k,z)\bar{x}_e(1-\bar{x}_e)\right] \, . \nonumber
\eea
The neutral Hydrogen density power spectrum is represented by $P_{g g}(k,z)$, $P_{\delta_x \delta_x}(k,z)$
is the power spectrum related to perturbations in the ionized fraction, $P_{g \delta_x}(k,z)$ is the cross-correlation power between
these two, and $P_{\delta_v \delta_v}(k,z)$ is the velocity divergence spectrum. Using linear theory
evolution of gravitational perturbations, the latter can be written as $P_{\delta_v \delta_v}(k,z) \approx f^2(\Omega_m) P^{\rm lin}(k,z)$
in terms of the linear density spectrum. The cross spectra between velocity and neutral density and
ionized fraction are given as $P_{g \delta_v}$ and $P_{\delta_v \delta_x}(k,z)$, respectively.

We assume the neutral Hydrogen density field is an unbiased tracer of the dark matter density field and write
$P_{g g}(k,z)=P_{\delta \delta}(k,z)$ and, when illustrating our results, 
use the linear theory spectrum to describe dark matter around the era of reionization. More complicated models that capture the non-linear
behavior of the gas density field can be obtained through approaches such as the halo models of large scale structure \cite{cooshe02}, 
but extended to higher redshifts. Note that the correlations in the ionized fraction depend on the reionization history and we
make use of the  patchy reionization model in Santos et al (2003) to write
\be                                                                                                        
\label{patchy}
P_{\delta_x \delta_x}(k,z)=b^2(z)P_{\delta \delta}(k,z) e^{-k^2 R^2}
\ee
where $b(z)$ is a mean bias that captures the halo bias \cite{Moetal97} 
weighted by the different halo properties, and $R$ the mean radius of the
HII patches. The time-dependence of $R$ is modeled as $R=\left({1\over 1-\bar{x}_e}\right)^{1/3} R_p$ 
where $R_p$ is the (comoving) size of the fundamental patch taken to
be $\sim 100$ Kpc.  

In addition to the power spectra of the gas density, velocity divergence, and ionization fraction,
when presenting numerical results, we assume cross-correlation spectra  between each of these three fields can be described with
perfect correlations such that for example, $P_{xy}=\sqrt{P_{xx}P_{yy}}$ where $x$ and $y$ are any of the two fields of interest.

In Fig.~1, we show the power spectrum of 21 cm fluctuations at a redshift of 10 with a bandwidth for observations of 1 MHz. Here, we make
use of a reionization model from Santos, Cooray \& Knox (2004), but similar results, with amplitudes varying as shown in Fig.~2 of
Santos, Cooray \& Knox (2004), are found for other models. The reionization model provides information on $\bar{x}_e$ and ionizing source bias.
In Fig.~1, we separate the contribution to the 21 cm anisotropy power spectrum to those related to the neutral density/ionizing fraction fields ($P_{aa}$) and peculiar velocities ($P_{bb}$). In addition to these two, the cross spectra between  these quantities ($P_{ab}$) 
also contribute to the total 21 cm anisotropy  angular power spectrum and this leads to a roughly a factor of 2 increase in power at large 
angular scales when compared to density fluctuations alone. 

In the bottom panel of Fig.~1, we show how this fraction varies as a function of
the observational bandwidth. In addition to the ratio of total to density contributions, we also show the fractional contribution related to
peculiar velocity fluctuations alone. For all bandwidths, this contribution is at the level around 0.3; one cannot simply
ignore the presence of peculiar velocities based on assumptions related to a large bandwidth for observations (c.f., Zaldarriaga, Furlanetto \&
Hernquist 2004). Note that the importance of peculiar velocities has already been discussed in Bharadwaj \& Ali (2004) in the
context of 21 cm fluctuations prior to reionization in the form of an absorption signal related to background CMB.


\begin{figure}
\centerline{\psfig{file=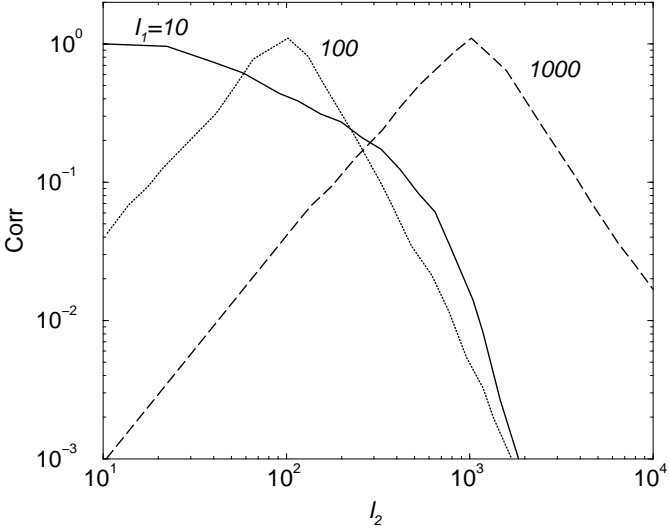,width=3.5in,angle=-90}}
\caption{The correlation coefficients related to the non-Gaussian covariance of 21 cm power spectrum measurements at $z=10$
with a bandwidth for observations of 1 MHz. Here, we show
${\rm Cov}^{\rm NG}_{l_1 l_2}/\sqrt{{\rm Cov}^{\rm NG}_{l_1 l_1} {\rm Cov}^{\rm NG}_{l_2 l_2}}$ as a function of $l_2$ when
$l_1=10$, 100, and 1000. Note that these are correlations only in the non-Gaussian part of the contribution to the
covariance. The total correlations, when considering the measurement of the 21 cm angular power spectrum, include the part due to
Gaussian variance. Since Gaussian variance is roughly two orders of magnitude or more higher than that of non-Gaussianities,
as shown in Fig.~1, these correlations are at most a percent and can be ignored for all practical purposes.}
\label{fig:corr}
\end{figure}

\subsection{Power spectrum covariance}

In prior works related to the 21 cm anisotropies, the fluctuations were assumed to be Gaussian so that the power
spectrum captures all statistical information and its variance is simply given by Gaussian statistics (e.g., Zaldarriaga, Furlanetto \& Hernquist 2004; Santos, Cooray \& Knox 2004). The second order
corrections to the 21 cm brightness temperature, in an era of partial reionization, however, leads to a non-Gaussian
contribution to the 21 cm brightness temperature anisotropies. These non-Gaussianities, then, add an extra covariance component to the
power spectrum measurement and a reliable estimate of the signal-to-noise ratio of the power
spectrum measurement should account for these, and other potentially interesting, non-Gaussian  effects. 

Writing the measured power spectrum as
\begin{equation}
\hat{C}_l = \frac{1}{2l+1} \sum_{m=-l}^l \langle a_{lm} a_{lm}^\star \rangle \, ,
\end{equation}
we note that the full covariance of this measured power spectrum is
\begin{eqnarray}
&&{\rm Cov}_{l_1 l_2} = \\
&&\frac{1}{2l_1+1} \frac{1}{2l_2 +1} \sum_{m_1 m_2} \langle a_{l_1 m_1} a_{l_1 m_1}^\star a_{l_2 m_2} a_{l_2 m_2}^\star \rangle - \hat{C}_{l_1} \hat{C}_{l_2} \, , \nonumber
\end{eqnarray}
and separate this to part which is Gaussian and a part which is non-Gaussian such that
\begin{eqnarray}
{\rm Cov}_{l_1 l_2} &=& {\rm Cov}^{\rm Gau}_{l_1 l_2} + {\rm Cov}^{\rm NG}_{l_1 l_2} \\
{\rm Cov}^{\rm Gau}_{l_1 l_2} &=& \frac{2}{2l_1+1} C_{l_1}^2 \delta_{l_1 l_2} \nonumber \\ 
{\rm Cov}^{\rm NG}_{l_1 l_2} &=& \frac{1}{2l_1+1} \frac{1}{2l_2 +1} \sum_{m_1 m_2} \langle a_{l_1 m_1} a_{l_1 m_1}^\star a_{l_2 m_2} a_{l_2 m_2}^\star \rangle_c \, , \nonumber
\end{eqnarray}
where $\langle .. \rangle_c$ denotes the connected part of the four-point correlator.
To calculate this non-Gaussian part, we now make use of the second order correction and, here, only focus on the part involving
the convolution between neutral Hydrogen density and ionization fraction fluctuations.
In terms of multipole moments, this second order term can be written as
\begin{eqnarray}
&&a^2_{l m} = -(4\pi)^2 \int \frac{d^3\bk_1}{(2\pi)^3} \int \frac{d^3\bk_2}{(2\pi)^3} \sum_{l_1 m_1} \sum_{l_2 m_2} (-i)^{l_1+l_2} \nonumber \\
&\times& \int d\rad c \bar{x}_e j_{l_1}(k_1 \rad) j_{l_2}(k_2 \rad) \delta_x(\bk_1) \delta_g(\bk_2) \\
&\times& Y_{l_1}^{m_1 \star}(\hat{\bf k}_1) Y_{l_2}^{m_2 \star}(\hat{\bf k}_2) \int d\hat{\bf n} Y_{l}^{m \star}(\hat{\bf n}) Y_{l_1}^{m_1}(\hat{\bf n}) Y_{l_2}^{m_2}(\hat{\bf n}) \, . \nonumber
\end{eqnarray}

Using this term twice, we simplify to obtain
\begin{eqnarray}
&&\langle a_{l_1 m_1} a_{l_1 m_1}^\star a_{l_2 m_2} a_{l_2 m_2}^\star \rangle_c = \\
&&\frac{8}{\pi^3} \int k_1^2 dk_1 \int k_2^2 dk_2 \int k_3^2 dk_3 c^4 \bar{x}_e^2 \nonumber \\
&\times&\Big[\bar{x}_e^2 P_{x_e x_e}(k_1) P_{x_e x_e}(k_2) P_{gg}(k_3) \nonumber \\
&& \quad + (1-\bar{x}_e)^2 P_{gg}(k_1) P_{gg}(k_2) P_{x_e x_e}(k_3) \Big] \nonumber \\
&\times& \sum_{m_1 m_2 l m} I_{l_2}(k_2) I_{l_1}(k_1) K_{l_1,l}(k_1,k_3) K_{l_2,l}(k_2,k_3) \nonumber \\
&\times& \int d\hat{\bf n} Y_{l_1}^{m_1}(\hat{\bf n}) Y_{l_2}^{m_2}(\hat{\bf n}) Y_{l}^{m}(\hat{\bf n}) \nonumber \\
&\times& \int d\hat{\bf m} Y_{l_2}^{m_2}(\hat{\bf m}) Y_{l_1}^{m_1}(\hat{\bf m}) Y_{l}^{m}(\hat{\bf m}) \, , \nonumber
\end{eqnarray}
where 
\begin{eqnarray}
K_{l_1,l_2}(k_1,k_2) &=& \int d\rad W_\nu(\rad) j_{l_1}(k_1\rad) j_{l_2}(k_2\rad) \, .
\end{eqnarray}
Including permutations, the non-Gaussian part of the power spectrum covariance is then
\begin{eqnarray}
&&{\rm Cov}_{l_1 l_2}^{\rm NG} = \\
&&\frac{8}{\pi^4} \sum_l (2l+1) \left(
\begin{array}{ccc}
l_1 & l_2 & l \\
0 & 0  &  0
\end{array}
\right)^2 
\int k_1^2 dk_1 \int k_2^2 dk_2 \nonumber \\
&\times& \int k_3^2 dk_3 c^4 \bar{x}_e^2 
\Big[\bar{x}_e^2 P_{x_e x_e}(k_1) P_{x_e x_e}(k_2) P_{gg}(k_3) \nonumber \\
&& \quad + (1-\bar{x}_e)^2 P_{gg}(k_1) P_{gg}(k_2) P_{x_e x_e}(k_3) \Big] \nonumber \\
&\times& I_{l_2}(k_2) I_{l_1}(k_1) K_{l_1,l}(k_1,k_3) K_{l_2,l}(k_2,k_3) \, , \nonumber
\end{eqnarray}
and an additional permutation with the replacement of $l_1 \rightarrow l_2$. 
Here, in simplifying, we have made use of the orthonormality relation between Wigner-3j symbols.
Note that we have ignored the cross power spectra between
density and ionization fraction here since these are likely to be subdominant relative to the density and ionization fraction power
spectra alone. Furthermore, we have ignored the non-Gaussian signals associated with the coupling between density fluctuations
and peculiar velocities. These are subdominant due to mismatch in projection between density and velocity fluctuations, captured
by projection integrals of $I_l$ with a $j_l$, and $J_l$ with a $j''_l$.

In Fig.~1, we show the variance related to Gaussian cosmic variance term and the non-Gaussian contribution separately
in the form of an error on the measured power spectrum and unbinned in multipole space. These errors are equivalent to that of
$\sqrt{{\rm Cov}_{l l}}$ but weighted by a factor of $l(l+1)/2\pi$. As shown, the non-Gaussian correction to the unbinned
power spectrum measurement error is two orders of magnitude below when compared to that due to Gaussian errors.

In Fig.~2, we show the correlation coefficient related to the non-Gaussian covariance where we plot
${\rm Corr} = {\rm Cov}^{\rm NG}_{l_1 l_2}/\sqrt{{\rm Cov}^{\rm NG}_{l_1 l_1} {\rm Cov}^{\rm NG}_{l_2 l_2}}$
as a function of $l_2$ for several values of $l_1$. While these correlations are large, note that what matters for the
power spectrum measurement is the total correlations, or rather,
${\rm Cov}^{\rm NG}_{l_1 l_2}/\sqrt{{\rm Cov}_{l_1 l_1} {\rm Cov}_{l_2 l_2}}$. Since ${\rm Cov}_{l_1 l_1} = {\rm Cov}_{l_1 l_1}^{\rm Gau}+{\rm Cov}_{l_1 l_1}^{\rm NG} \approx {\rm Cov}_{l_1 l_1}^{\rm Gau}$, these correlations are below a percent at most (when $l \sim$ few $10^3$)
and can be ignored for all reasonable purposes involving the measurement of the 21 cm power spectrum.

\begin{figure}
\centerline{\psfig{file=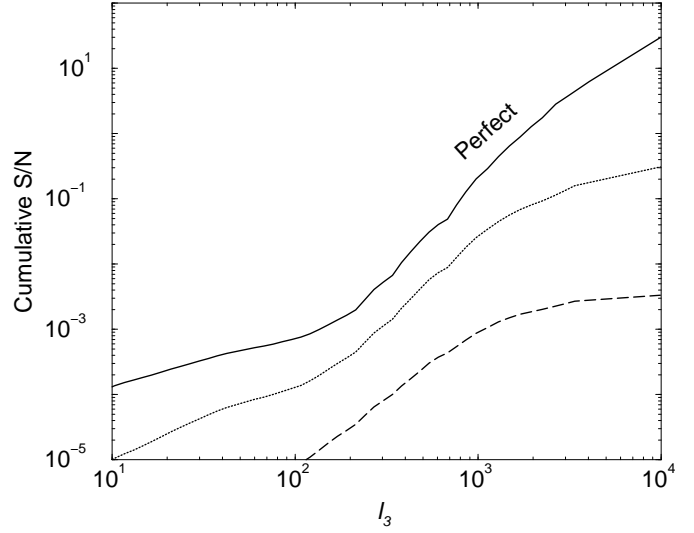,width=3.5in,angle=-90}}
\caption{Cumulative signal-to-noise ratio for the measurement of the 21 cm bispectrum as a function of $l_3$, for
observations at $z=10$ with a bandwidth of 1 MHz. From top-to-bottom, the
three curves for an all-sky experiment with no instrumental noise, a 4000 deg.$^2$ survey with instrumental noise as show in
Fig.~1, and a survey of 400 deg.$^2$ and with an increase in noise by a factor of 2 to account for unsubtracted foregrounds
following Santos, Cooray \& Knox (2004). While the signal-to-noise ratio is at the level of a few tens for the {\it perfect}
experiment, for more realistic scenarios, the signal-to-noise ratio is below a unity.}
\label{fig:bispec}
\end{figure}

\subsection{Bispectrum}

While corrections to the power spectrum covariance from non-Gaussianities are negligible, it is still useful to consider
if non-Gaussianities in the 21 cm background can be measured with upcoming interferometers.
To calculate the non-Gaussian signal, here, we concentrate on the lowest order statistic 
and estimate the signal-to-noise ratio associated with the
bispectrum, or the Fourier space analogue of the three-point correlation function. 
For reference,  the three-point correlation function is related to multipole moments via
\begin{eqnarray}
B(\bn,\bm,\bl) &\equiv& \langle T(\bn)T(\bm)T(\bl) \rangle \\
               &\equiv&
                \sum 
                \langle \alm{1} \alm{2} \alm{3} \rangle
                \Ylm{1}(\bn) \Ylm{2}(\bm)  \Ylm{3}(\bl)\,,\nonumber
\end{eqnarray}
where the sum is over $(l_1,m_1),(l_2,m_2),(l_3,m_3)$.
Statistical isotropy allows us to express the correlation in terms an $m$-independent function,
\begin{eqnarray}
\langle \alm{1} \alm{2} \alm{3} \rangle  = \wjm \bi\,,
\end{eqnarray}
where the quantity $\bi$ is described as the angular averaged bispectrum \cite{Cooray:1999kg}
and we have made use of the orthonormality relation related to Wigner-3j symbols.
As discussed in Cooray \& Hu (2000), the angular bispectrum, $\bi$, contains all the information available
in the three-point correlation function and frequently used quantities such as the skewness and the collapsed three-point function
can be written in terms of the bispectrum, such as a filtered sum of certain bispectrum configurations \cite{Cooray:2000uu}.

We consider the bispectrum formed by the 21 cm brightness fluctuations. This non-Gaussianity arises from the
second order terms in Eq.~3 (or in Fourier space, Eq.~9). Here, we again ignore the correction related to peculiar velocities, 
but instead focus on the second order term 
related to the coupling between neutral gas density  fluctuations
and the ionization fraction distribution. 
The bispectrum is simply then
\begin{eqnarray}
&&\langle a^2_{l_1 m_1}a_{l_2 m_2}a_{l_3 m_3} \rangle = \nonumber \\
&& {4 \over \pi^2} \int k_1^2 dk_1 \int k_2^2 dk_2 c^3 (1-\bar{x}_e) \bar{x}_e^2 I_{l_2}(k_1) I_{l_3}(k_2) \nonumber \\
&\times& K_{l_2,l_3}(k_1,k_2) \left[P_{g g}(k_1) P_{x_e x_e}(k_2) + P_{g x_e}(k_1) P_{g x_e}(k_2) \right] \nonumber \\
&\times& \int d\hat{\bf n} Y_{l_1}^{m_1}(\hat{\bf n}) Y_{l_2}^{m_2}(\hat{\bf n}) Y_{l_3}^{m_3}(\hat{\bf n}) \, ,
\label{eqn:bispec}
\end{eqnarray}
where, again, we have made use of the $a^2_{l m}$ term from Eq.~19.

The bispectrum is given by
\begin{eqnarray}
B_{l_1 l_2 l_3} &=& \sum_{m_1 m_2 m_3} \wjm \left[
\left< a^2_{l_1 m_1}a_{l_2 m_2}a_{l_3 m_3}  \right> + {\rm Perm.} \right]\nonumber \\
&=& \sqrt{\frac{(2l_1 +1)(2 l_2+1)(2l_3+1)}{4 \pi}}
\left(
\begin{array}{ccc}
l_1 & l_2 & l_3 \\
0 & 0  &  0
\end{array}
\right) \nonumber \\
 && \quad \quad \quad \times \left[ b_{l_2,l_3} + {\rm Perm.} \right]\, . 
\label{eqn:ovbidefn}
\end{eqnarray}
Here,
\begin{eqnarray}
 b_{l_2,l_3}
&=& {4 \over \pi^2} \int k_1^2 dk_1 \int k_2^2 dk_2 c^3 (1-\bar{x}_e) \bar{x}_e^2 \nonumber \\
&&\times\left[P_{g g}(k_1) P_{x_e x_e}(k_2) + P_{g x_e}(k_1) P_{g x_e}(k_2) \right] \nonumber \\
&&\times I_{l_2}(k_1) I_{l_3}(k_2) K_{l_2,l_3}(k_1,k_2)   \, .
\label{eqn:finalintegral}
\end{eqnarray}
The permutations here account for the 3 additional terms with the replacement of $l_2 \rightarrow l_3$ and $l_1$.

The signal-to-noise ratio for the bispectrum is calculated as
\begin{equation}
\left(\frac{{\rm S}}{{\rm N}}\right)^2 = f_{\rm sky} \sum_{l_3 \geq l_2 \geq l_1}
        \frac{\left(B_{l_1 l_2 l_3}\right)^2}{
          C_{l_1}^{\rm tot}
          C_{l_2}^{\rm tot}
          C_{l_3}^{\rm tot}}\,,
\label{eqn:chisq}
\end{equation}
Here, $C_l^{tot}$ represents all contributions to the power spectrum of the $i$th field,
\begin{eqnarray}
C_l^{tot} = C_l + C_l^{\rm noise} + C_l^{\rm foreg}\, ,
\label{eqn:cltot}
\end{eqnarray}
where $C_l^{\rm noise}$ is the instrumental noise contribution related to a 21 cm observations, 
and $C_l^{\rm foreg}$ is the confusing foreground contribution. The expression assumes all-sky  observations of the 21 cm 
background, but in the case of
partial sky observations, the square of the signal-to-noise ratio is reduced
by the fraction of sky covered by the data, $f_{\rm sky}$.
We make use of the noise spectrum shown in Fig.~1 and include foregrounds in terms of the
residual noise left over after multi-frequency cleaning following Santos, Cooray \& Knox (2004).
This procedure generally results in a factor of $\sim$ 2 increase in instrumental noise
in the mutlipole range of general interest.

In Fig.~3, we summarize our results related to the signal-to-noise ratio calculation. While the signal-to-noise
ratio is at the level of a few tens for a no instrumental noise experiment that map out essentially the all
sky, for partial sky-coverages and for added instrumental and foreground noise, the signal-to-noise
ratio decreases below a unity suggesting that a reliable detection of the 21 cm anisotropy bispectrum
may not be possible with first-generation low-frequency radio interferometers.

\section{Discussion \& Summary}
\label{discussion}

The brightness temperature fluctuations in the 21 cm background related to neutral Hydrogen content
provide a useful probe of physics related to the era of reionization during the transition from a
completely neutral intergalactic medium to a partially ionized one. Here, we have presented a complete
description of brightness temperature anisotropies related to 21 cm background  in terms of fluctuations in the
spatial distribution of density, peculiar velocity and ionization-fraction. Our calculations are
more appropriate to the large physical, and thus large angular, scales since we have not attempted to
model the small-scale non-linear behavior of the reionization process as well as effects associated with
finite size of ionizing patches, among others. 

Our formalism allows us to calculate the  angular power
spectrum, the bispectrum, the Fourier analog of the three-point correlation function, and the trispectrum,
the Fourier analog of the four-point correlation function, of 21 cm fluctuations. In the case of the power spectrum,
we have shown that peculiar velocities cannot easily be ignored since variations in the large scale velocity field
and its cross-correlation with the density distribution lead to an additional contribution to the angular
power spectrum of 21 cm anisotropies. For simple models related to the correlation between density
and velocity fluctuations, the correction is at the level of a factor of 2. As shown in Fig.~1, at large angular scales,
this correction is independent of the bandwidth of 21 cm observations while with increasing bandwidth, the correction
reduces at arcminute angular scales corresponding to multipoles of a 10$^3$ and higher.

We also extend previous discussions related to the power  spectrum measurement by including an estimate of the non-Gaussian
covariance resulting from parallelogram configurations of the trispectrum. In the case of the variance, in
terms of an unbinned error on the power spectrum, the non-Gaussian correction is at least two orders of magnitude at
degree angular scales with an increase to slightly less than an order of magnitude at multipoles of 10$^4$ (Fig.~1).
In the case of the covariance, in terms of correlations between power spectrum measurements at
different multipoles, the correlations are at most at the level of a few percent, when $l \sim 10^4$, suggesting that
for most practical purposes non-Gaussianities can be safely ignored. While the effect of non-Gaussianities on the
power spectrum measurement is insignificant, in terms of a direct measurement of the non-Gaussian signal,
we have explore the possibility related to a measurement of the bispectrum. As shown in Fig.~3, the signal-to-noise
ratio for the full bispectrum measurement is at most at the level of a few tens. For more practical
observations with the first generation experiments, this ratio drops below unity both due to added instrumental
noise and the small sky-coverage. It is unlikely that the non-Gaussian signal will be directly measurable
with 21 cm data alone. It could be possible to pull out additional information, especially related to the
non-linear mode coupling second order terms, using cross-correlations between small scale CMB anisotropies
(Cooray 2004b).

Thus, for most practical purposes, the presence of non-Gaussianities can be ignored and the 21 cm fluctuations can be 
described with Gaussian statistics. Since 21 cm fluctuations are significantly contaminated with foregrounds, such
as galactic synchrotron or low-frequency radio point sources, the lack of a significant non-Gaussianity in the
signal suggests that any significant detection of non-Gaussianity can be easily ascribed to a foreground confusion.
Similarly, in addition to frequency information that is now proposed to separate 21 cm fluctuations from foregrounds,
if non-Gaussian structure of foregrounds is a priori known, one can potentially use this additional information to
further reduce the foreground confusion. This requires a better understanding of the foreground distribution, especially
involving the non-Gaussian nature of foregrounds. While in Santos, Cooray \& Knox (2004), foregrounds were described
only in terms of the power spectra, the formalism can be modified to include this additional information and we hope to return
to this topic in a later discussion.

While our calculations show that non-Gaussianities in the 21 cm background may not be significant, we may have underestimated the
non-Gaussian nature by only concentrating on the large physical, and thus large angular, scales of the signal. For example,
additional non-Gaussianities are expected when the density field becomes non-linear. This, for redshifts $z \sim 10$,
happens at scales below the typical resolution of upcoming instruments and can be safely ignored. Similarly, the non-linear
nature of the reionization process, especially the finite patch size of reionization patches, could lead to an additional
non-Gaussian signal. This again depends on model dependent quantities and are better addresses with numerical simulations rather
than analytical calculations.

{\it Acknowledgments:}
Author acknowledges support from the Sherman Fairchild foundation and NASA NAG5-11985.

\end{document}